\documentclass[a4paper,12pt]{revtex4}
\usepackage{epsfig}

\def\be{\begin{equation}}
\def\ee{\end{equation}}
\def\bea{\begin{eqnarray}}
\def\eea{\end{eqnarray}}

\begin{document}

\title{Evidence for the higher twists effects in diffractive DIS at HERA}

\author{M.~Sadzikowski\,\footnote{Talk presented during the conference Rencontres de Moriond, 
"QCD and High Energy Interactions", 2012.}, L.~Motyka, W.~S\l{}omi\'{n}ski}

\affiliation{Smoluchowski Institute of Physics, Jagiellonian
University, Reymonta 4, 30-059 Krak\'ow, Poland}

\begin{abstract}
We study a twist decomposition of diffractive structure functions in the diffractive deep inelastic scattering at HERA.
At low $Q^2$ and at large energy the data exhibit a strong deviation from the twist-2 NLO DGLAP description.
It is found that this deviation in consistent with higher twist effects.
We conclude that the DDIS at HERA provides the first, strong evidence of higher twist effects in DIS.
\end{abstract}

%\pacs{74.20.Mn,03.65.Pm,}

\maketitle

\section{Introduction}

The QCD description of the diffractive deep inelastic scattering processes $ ep\rightarrow epX$ (DDIS) is based on the series expansion of 
the scattering amplitudes in the inverse powers of a large scale $Q^2$, defined as a negative squared four-momentum transfer from the electron to the proton
carried by the virtual photon $\gamma^\ast$. In the leading twist-2 approximation the diffractive proton structure functions
$F_{L,T}^{D(3)}$ can be calculated using diffractive parton distribution functions (DPDFs) due to the Collins factorization theorem
\cite{collins}, whereas the DPDFs dependence on the hard scale is governed by the celebrated DGLAP evolution equation. Despite of great
efficiency of this approximation in the data description this approach has an obvious limitation that follows from negligence
of the higher twists contributions. Certainly, the higher twists contribute at any energy scale and become relevant for data description
below some virtuality $Q^2$, which depends on the process and required precision. In this presentation we point out that
in the case of DDIS the DGLAP description breaks down at the scale $Q^2\simeq 5$ GeV$^2$ and to show that these deviations are consistent with 
a higher twists contribution. 

\section{Cross section and the DGLAP description}

\begin{figure}
\centerline{
\includegraphics[width=0.4\columnwidth]{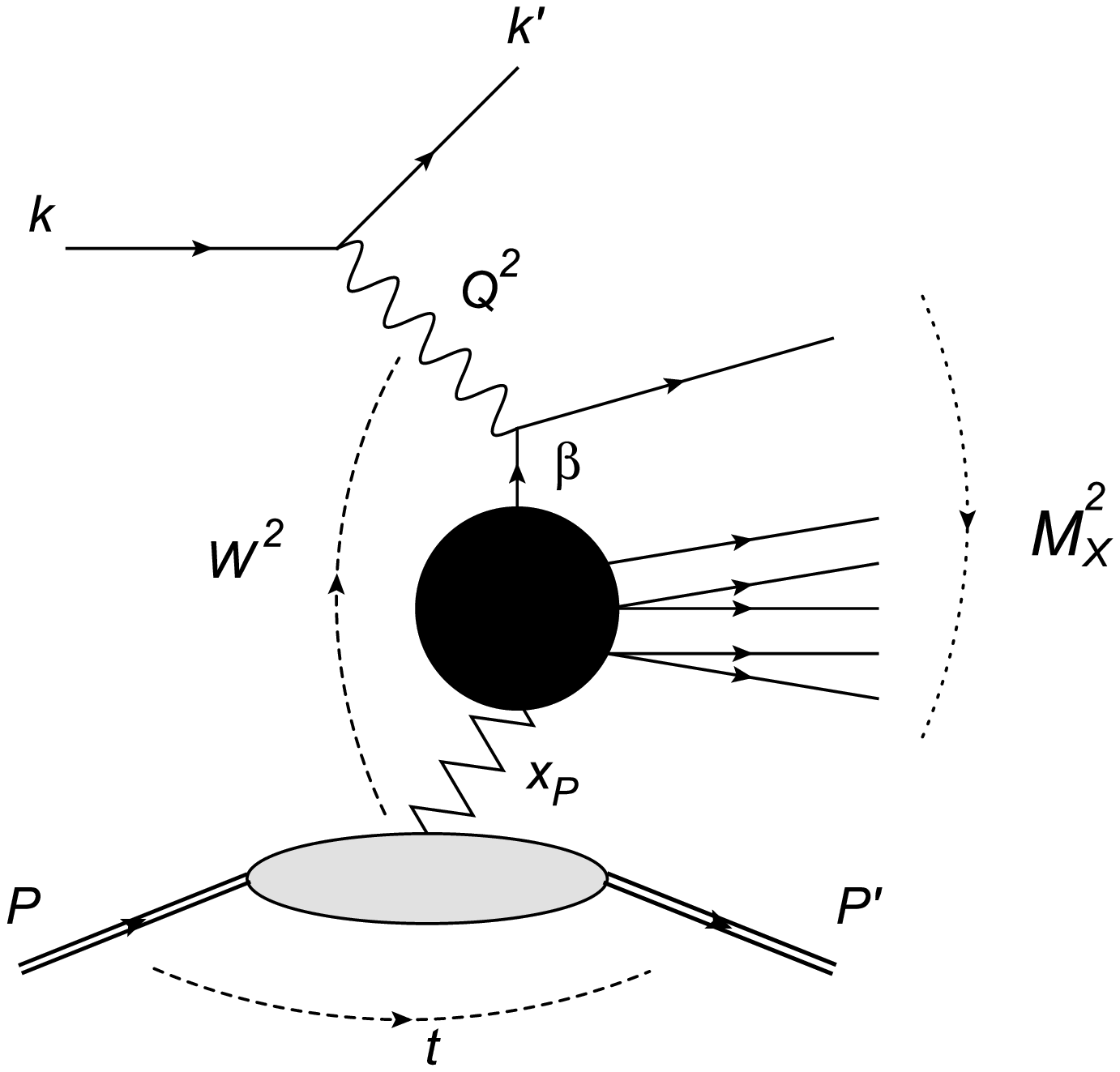}\vspace{-1em}\hspace*{1cm}\includegraphics[width=0.4\columnwidth]{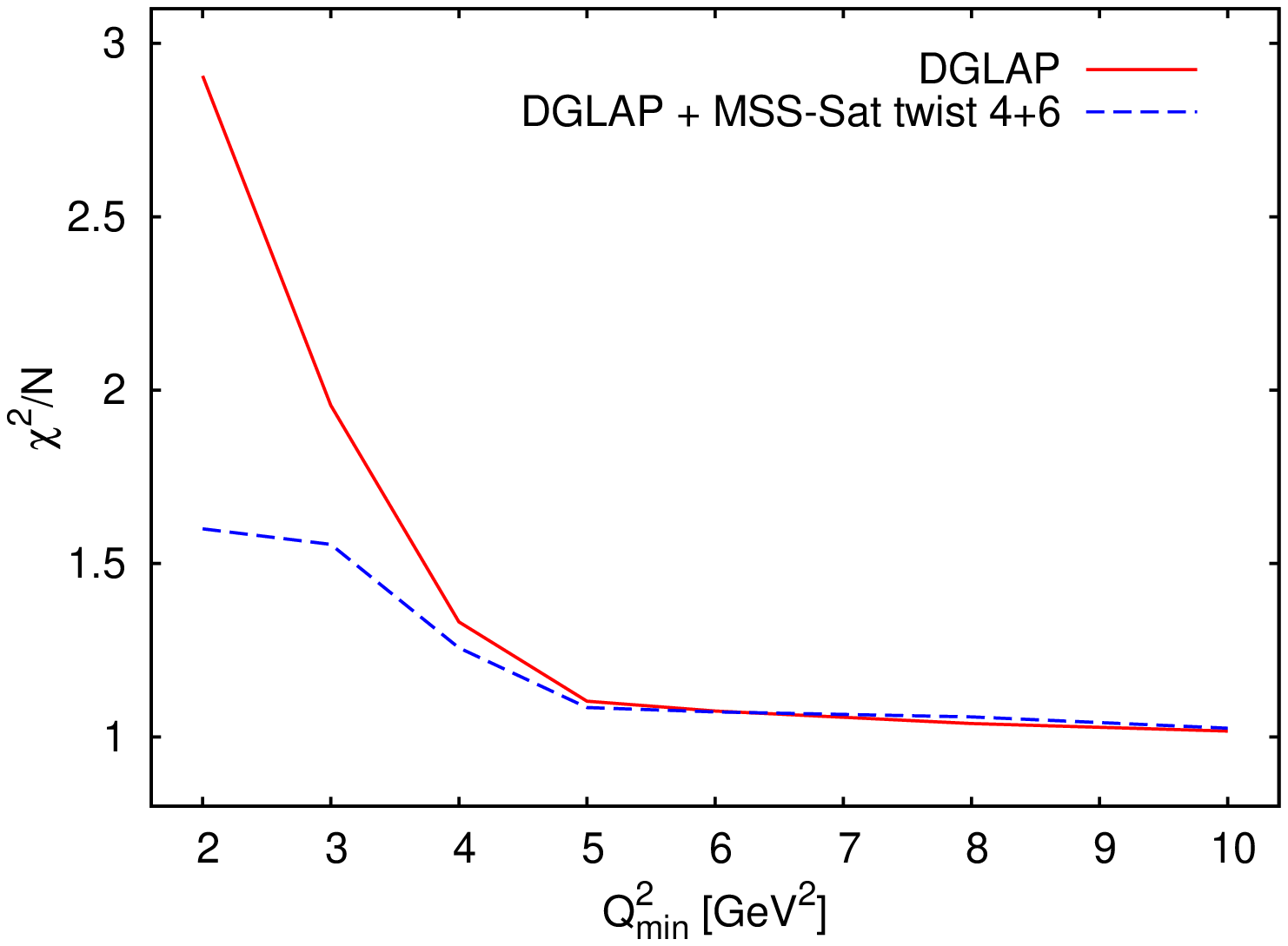}\vspace{-1em}
}
\vspace*{3em}
Fig. 1. Left panel - kinematics of the DDIS scattering. Right panel - the $\chi^2/\,{\rm d.o.f.}$ for NLO DGLAP and NLO DGLAP + HT fits to ZEUS LRG data \cite{ZEUS} with $Q^2 < Q^2_{\mathrm{min}}$.
%\caption{Left panel - kinematics of the DDIS scattering. Right panel - the $\chi^2/\,{\rm d.o.f.}$ for NLO DGLAP and NLO DGLAP + HT fits to ZEUS LRG data \cite{ZEUS} with $Q^2 < Q^2_{\mathrm{min}}$.
%\label{fig1}
%}
\end{figure}
The DDIS is an quasi-elastic electron-proton scattering process $e(k)p(P)\rightarrow e(k^\prime)p(P^\prime)X(P_X)$ in which the
final hadronic state $X$ with four-momentum $P_X$ is separated in rapidity from the proton, that scatters elastically (see Fig. 1).
The $t$-integrated $ep$ cross-section reads:
\be
\frac{d\sigma}{d\beta dQ^2 d\xi} = \frac{2\pi\alpha^2_{\mathrm{em}}}{\beta Q^4}[1+(1-y)^2]\sigma_r^{D(3)}(\beta, Q^2,\xi)
\ee
where the invariants read $y=(kq)/(kP)$, $Q^2=-q^2$, $\xi = (Q^2+M_X^2)/(W^2+Q^2)$ and $t=(P^\prime - P)^2$. The
quantity $W^2=(P+q)^2$ is the invariant mass squared in photon-proton scattering, and $M_X^2$ is the invariant mass
of the hadronic state $X$. The reduced-cross-section may be expressed in terms of the diffractive structure functions
\be
\sigma_r^{D(3)}(\beta, Q^2,\xi)=F_T^{D(3)}+ \frac{2-2y}{1+(1-y)^2}F_L^{D(3)},
\ee
whereas the structure functions $T,L$ may be, respectively, expressed through transversally and longitudinally polarized $\gamma^\ast$ - proton cross sections
$F_{L,T}^{D(3)} = (Q^4/4\pi^2\alpha_{\mathrm{em}}\beta\xi)d\sigma^{\gamma^\ast p}_{L,T}/dM^2_X$.

In the recent analysis \cite{ZEUS} the ZEUS diffractive data were fitted within NLO DGLAP approximation. A satisfactory
description was found only for $Q^2>Q^2_{\mathrm{min}}=5$ GeV$^2$. The ZEUS fits were performed
above $Q^2_{\mathrm{min}}$ and then extrapolated to lower photon virtualities. The deviations of the fits rapidly grow with
decreasing $\xi$ and $Q^2$ reaching 100 percent effect at the minimal $Q^2=2.5$ GeV$^2$ and $\xi\simeq 4\cdot 10^{-4}$.
We confirmed this result throug the calculation of $\chi^2/$d.o.f. for subsets of ZEUS LRG data with $Q^2>Q^2_{\mathrm{min}}$ and
$\beta > 0.035$ \cite{MSS} (see Fig. 1, right panel). The cut-off in $\beta$ is imposed to reject part of the data with significant
contributions from higher Fock states not included in our model.
It is clear from this discussion that the leading twist DGLAP evolution is unable to describe the DDIS data below
$Q^2\simeq 5$ GeV$^2$ and at the low $\xi$.

\section{Estimation of the higher twist contributions}

\begin{figure}
\centerline{
\includegraphics[width=0.4\columnwidth]{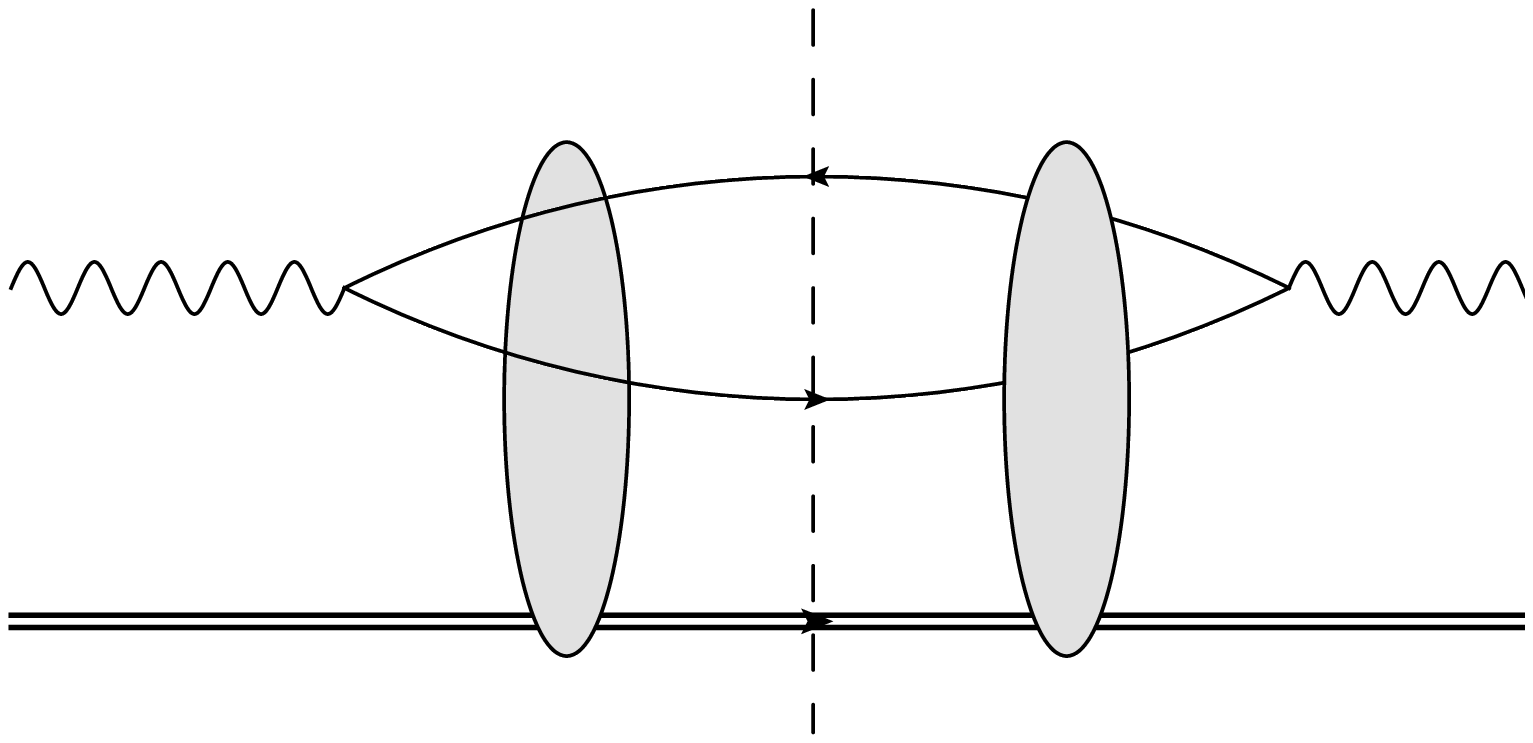}\vspace{-1em}\hspace*{1cm}\includegraphics[width=0.4\columnwidth]{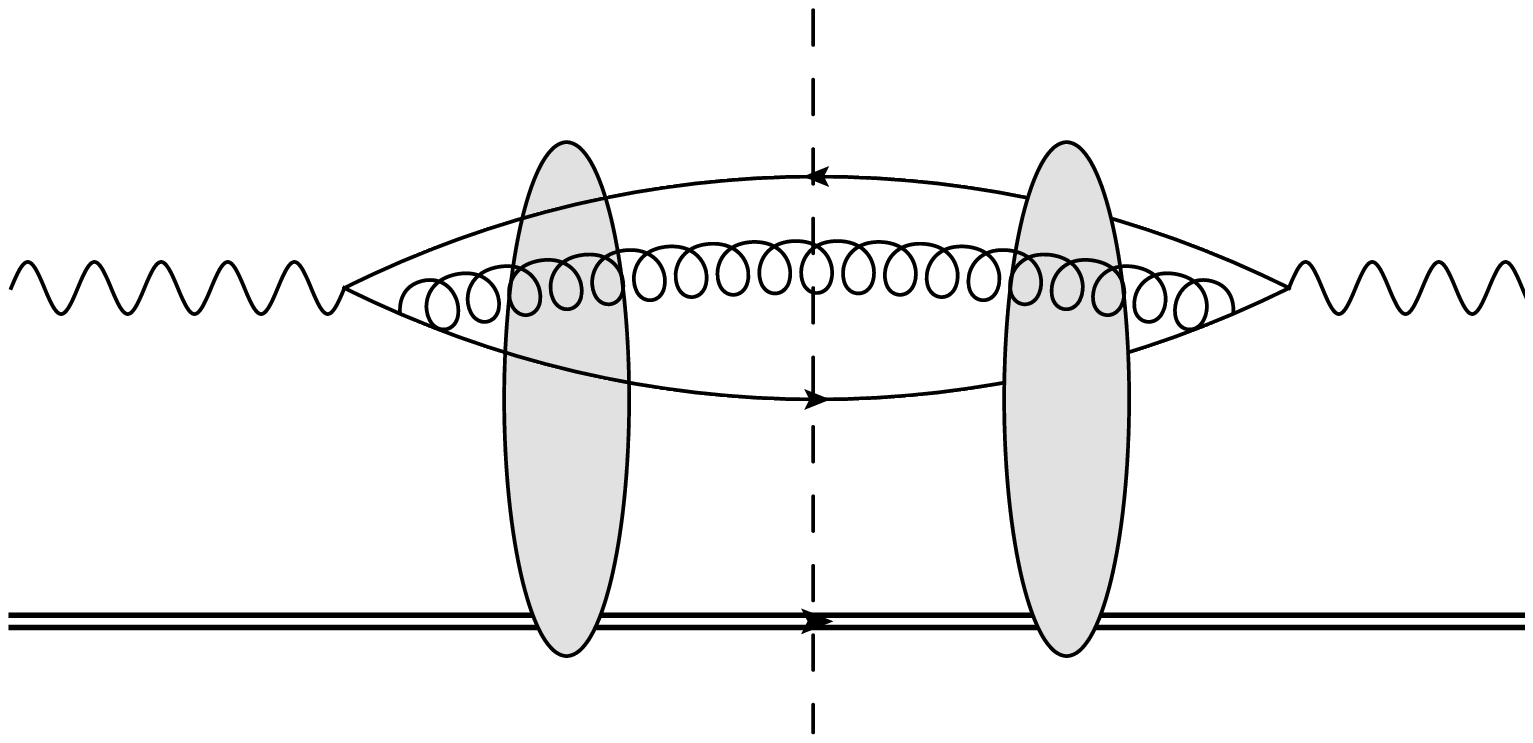}\vspace{-1em}
}
\vspace*{3em}
Fig. 2. Left panel - the quark box contribution. Right panel - the $q\bar{q}g$ contribution.
%\label{fig2}
%}
\end{figure}

The large energy limit of the DDIS scattering may be described within the framework of the
colour dipole model \cite{NZ,GBW}. In this approach the $\gamma^\ast p$ process is factorized
into an amplitude of photon fluctuation into the partonic debris and then scattering of these
states off the proton by the multiple gluon exchange. We take into account the contributions
from the fluctuation of the photon into a colour singlet quark-antiquark pair $q\bar{q}$ and
into $q\bar{q}$-gluon triple (see Fig. 2). This gives the $t$-integrated $\gamma^\ast p$ cross section
$d\sigma_{L,T}^{\gamma^\ast p}/dM_X^2 = d\sigma_{L,T}^{q\bar{q}}/dM_X^2+d\sigma_{L,T}^{q\bar{q}g}/dM_X^2$.

Assuming an exponential $t$-dependence of diffractive cross-section, one finds for the $q\bar{q}$ component (see Fig.2, left panel)
\be
\label{diff_cross_sec_qq}
\frac{d\sigma_{L,T}^{q\bar{q}}}{dM_X^2} = \frac{1}{16\pi b_D}\int\frac{d^2p}{(2\pi)^2}\int_0^1 dz \delta\left(\frac{p^2}{z\bar{z}}-M_x^2\right)
\sum_f\sum_{spin}\left| \int d^2r e^{i\vec{p}\cdot\vec{r}}\psi^f_{h\bar{h},\lambda}(Q,z,\vec{r})\sigma_d (r,\xi)\right|^2 .
\ee
where $b_D$ is a diffractive slope, $z\bar{z}=z(1-z)$ and the first sum runs over the three light flavours. The second sum of (\ref{diff_cross_sec_qq}) means 
summation over massless (anti)quark helicities $(\bar{h}) h$ in the case of longitudinal photons whereas for transverse photons there is an additional average over initial photon polarizations $\lambda$. The squared photon wave functions can be found in literature\cite{MKW}.

We use the GBW parametrization
\cite{GBW} for the dipole-proton cross section $\sigma_d(r,\xi) = \sigma_0(1-\exp(-r^2/4R_\xi^2))$ where the saturation radius in DDIS 
$R_\xi=(\xi /x_0)^{\lambda /2}$ GeV$^{-1}$ and $\sigma_0=23.03$ mb, $\lambda = 0.288$, $x_0=3.04\cdot 10^{-4}$.
The contribution of the $q\bar{q}g$ component of $\gamma^\ast$ (see Fig. 2, the right panel) is calculated at $\beta =0$ and in the soft gluon approximation (the
longitudinal momentum carried by the gluon is much lower then carried by the $q\bar{q}$ pair). This approximation
is valid in the crucial region of $M_X^2\gg Q^2$ or $\beta \ll 1$, where the deviations from DGLAP are observed.
The correct $\beta$-dependence is then restored using a method described by Marquet \cite{marquet}, with kinematically
accurate calculations of W\"sthoff \cite{wusthoff}. With these approximations one obtains:
\bea
\label{cross_sec_qqg}
\frac{d\sigma^{q\bar{q}g}_{L,T}}{dM_x^2} &=& \frac{1}{16\pi b_D}\frac{N_c\alpha_s}{2\pi^2}\frac{\sigma_0^2}{M_x^2}\int d^2r_{01}
N^2_{q\bar{q}g} (r_{01},\xi)\sum_f\sum_{spin}\int_0^1 dz|\psi^f_{h\bar{h},\lambda}(Q,z,r_{01})|^2, \\\nonumber
N^2_{q\bar{q}g} (r_{01}) &=& \int d^2r_{02}\frac{r_{01}^2}{r_{02}^2 r_{12}^2}\left( N_{02}+N_{12}-N_{02}N_{12}-N_{01} \right)^2
\eea
where $N_{ij} = N(\vec{r}_j-\vec{r}_i)$, $\vec{r}_{01}, \vec{r}_{02},\vec{r}_{12}=\vec{r}_{02}-\vec{r}_{01}$ denote the relative positions of quark and antiquark $(01)$, quark and gluon $(02)$ in the transverse plain. The form of $N^2_{qqg}$ follows from the Good-Walker picture of the diffractive dissociation of the photon \cite{munier_shoshi}. The factor $1/M_X^2$ is a remnant of the phase space integration under the soft gluon assumption.
The twist decomposition of (\ref{diff_cross_sec_qq}) is performed through the Taylor expansion in the inverse powers of $QR$ whereas
that of (\ref{cross_sec_qqg}) using Mellin transform technic \cite{MSS}.

\section{Discussion}

\begin{figure}
\centerline{
\includegraphics[width=0.45\columnwidth]{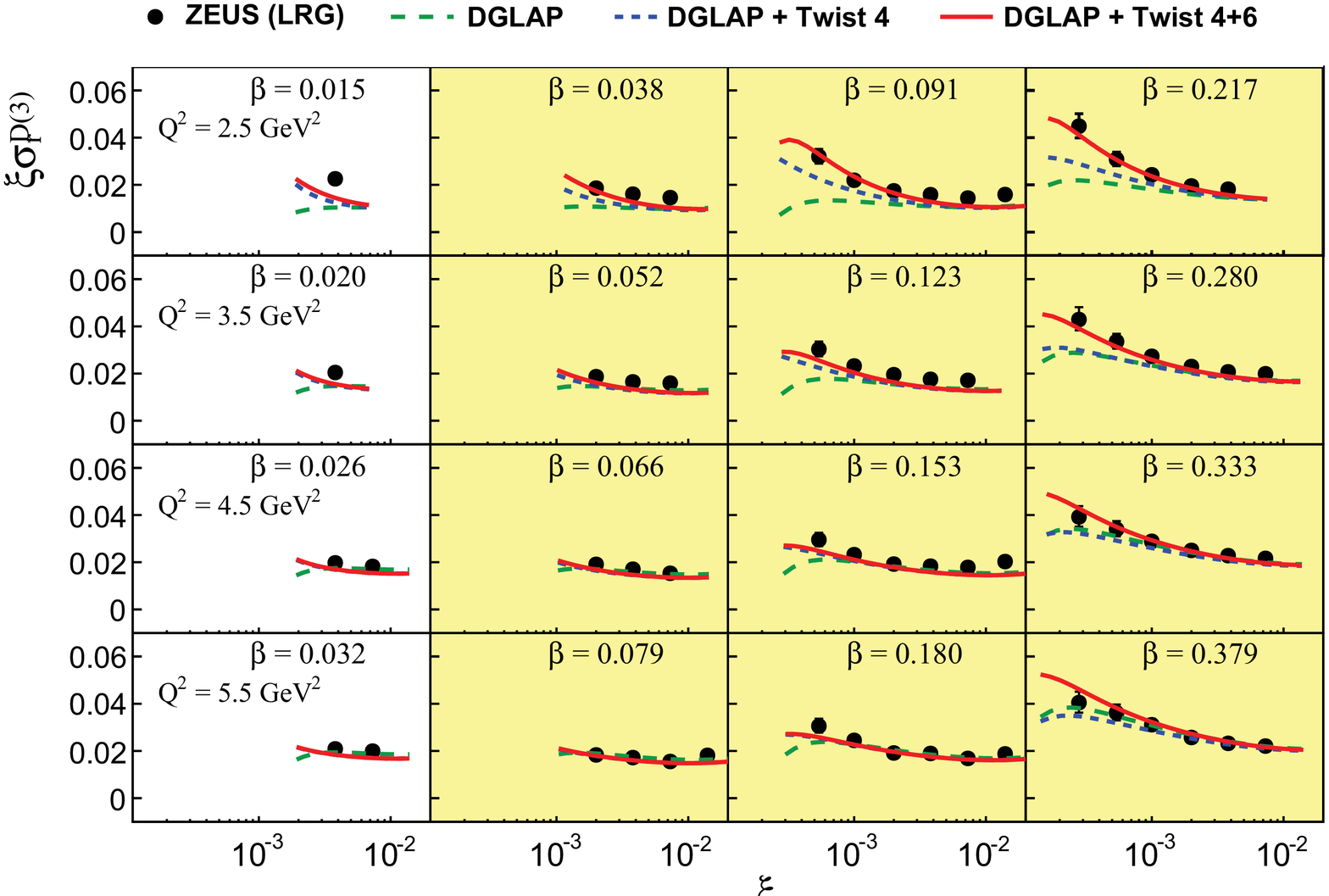}\vspace{-1em}
}
\vspace*{3em}
Fig. 3. The LRG ZEUS data for $\xi\sigma_r ^{D(3)}$ at low $Q^2$ compared to
%left panel - a DGLAP fit and twist-2 component of the saturation model. Right panel -  
a DGLAP fit \cite{ZEUS} and the DGLAP fit with included twist-4 and twist-4 and~6 corrections from the MSS saturation model. In yellow (gray) --- the region of $\beta$ where the correction due to $q\bar q gg$ may be neglected.
%\label{fig3}
%}
\end{figure}
In Fig. 3 we compare selected results with data.
 The saturation model (MSS model) results are obtained using the original GBW parameters $\lambda$ and $\sigma_0$, and three massless quark
flavours. In our approach we modified the GBW parameter $x_0$ to $\xi_0=2x_0$ in order to account for the difference between
Bjorken $x$ and pomeron $\xi$, the variables used in GBW dipole cross-section in DIS and DDIS respectively. We chose $\alpha_s=0.4$
that provides a good description of the data. The conclusion from the analysis and from Fig. 3 is that a combination of the DGLAP fit
and twist-4 and twist-6 components of the model gives a good description of the data at low $Q^2$.
%The conclusions from the analysis and from the Fig. 3
%are the following: (i) at twist-2 the DGLAP fit extrapolation, and the twist-2 components of the model agree with one another, but all fail to describe the data below $Q^2=5$ GeV$^2$ and at low %$\xi$ (left panel); (ii) a combination of the DGLAP fit
%and twist-4 and twist-6 components of the model gives a good description of the data at low $Q^2$ (right panel). 
Inclusion of these higher twist terms improves the fit quality in the low $Q^2$ region (see the dashed curve at Fig. 1 right panel).
Indeed, the maximal value of $\chi^2/$d.o.f. $\simeq 1.5$ at $Q^2_{min}=2$ GeV$^2$ is significantly lower then  $\chi^2/$d.o.f.
$\simeq 3$ of the DGLAP fit. Nevertheless, it is important to stress that a truncation of the twist series (up to twist-6) is required
to have a good description of the data. The truncation of this kind, however, may be motivated in QCD. Let
us recall that in BFKL, at the leading logarithmic approximation, only one reggeized gluon may couple to a fundamental
colour line. Since DGLAP and BFKL approximations have the same double logarithmic ($\ln x \ln Q^2$) limit, one concludes that also in DGLAP couplings of more than two gluons to a colour dipole is much weaker than in the eikonal picture. Thus one can
couple only two gluons to a colour dipole and up to four gluons to $q\bar{q}g$ component (two colour dipoles in the large $N_c$ limit)
without BFKL constraint. This means that one may expect a suppression beyond twist-8 if only the $q\bar{q}$ and $q\bar{q}g$
components are included in the calculations.

In conclusion, the DDIS data at low $Q^2$ provide the first evidence for higher twists effects in DIS in the perturbative domain
and opens a possibility for further theoretical and experimental investigations.

\textbf{Acknowledgments}
The work is supported by the Polish National Science Centre grant no. DEC-2011/01/B/ST2/03643.

\end{document}